\documentclass[aps,letterpaper,pra,twocolumn,floats,floatfix]{revtex4}
\usepackage{amsmath}

\usepackage{ifthen}
\usepackage{ifpdf}

\usepackage{graphicx}
\usepackage{color}
\usepackage{bm}
\usepackage{latexsym}
\usepackage{hyperref}

\ifpdf
\usepackage{graphicx}
\usepackage{epstopdf}
\else
\usepackage{graphicx}
\usepackage{epsfig}
\fi

\newcommand{\ket}[1]{\left|#1\right\rangle }
\newcommand{\mass}{\mathsf{m}}

\newcommand{\be}[1]{\begin{eqnarray}\ifthenelse{#1=-1}{\nonumber}{\ifthenelse{#1
=0}{}{\label{e#1}}}}
\newcommand{\ee}{\end{eqnarray}}

\newcommand{\hide}[1]{}
\newcommand{\tbox}[1]{\mbox{\tiny #1}}

\newcommand{\amatrix}[1]{\begin{matrix} #1 \end{matrix}}

\newcommand{\mcal}[1]{\mathcal{#1}}

\newcommand{\dog}[1]{#1^{\dagger}}
\newcommand{\scap}[2]{\langle#1|#2\rangle}
\newcommand{\braket}[3]{\langle#1|#2|#3\rangle}

\graphicspath{{epsfigs/}{./}}
\begin{document}

\normalsize

\title{Controlled quantum stirring of Bose-Einstein condensates}
\author{Moritz Hiller,$^{1,2,3}$ Tsampikos Kottos,$^{1}$ and Doron Cohen$^4$
}

\affiliation{
$^1$Department of Physics, Wesleyan University, Middletown, Connecticut 06459, USA \\
$^2$MPI for Dynamics and Self-Organization, Bunsenstra\ss e 10, D-37073 G\"ottingen, Germany \\
$^3$Physikalisches Institut, Albert-Ludwigs-Universit\"at, Hermann-Herder-Str.~3, D-79104 Freiburg, Germany\\
$^4$Department of Physics, Ben-Gurion University, Beer-Sheva 84105, Israel
}

\begin{abstract}
By cyclic adiabatic change of two control parameters 
of an optical trap one can induce a circulating 
current of condensed bosons. The amount of particles 
that are transported per period depends on the ``radius" 
of the cycle, and this dependence can be utilized 
in order to probe the interatomic interactions. 
For strong repulsive interaction the current can 
be regarded as arising from a sequence of Landau-Zener crossings. 
For weaker interaction one observes either gradual 
or coherent mega crossings, while for attractive 
interaction the particles are glued together and behave 
like a classical ball. For the analysis we use 
the Kubo approach to quantum pumping with the associated 
Dirac monopoles picture of parameter space.   
\end{abstract}

\pacs{03.65.-w, 03.65.Vf, 03.75.Lm, 05.30.Jp, 05.45.-a}
\maketitle


\section{Introduction}

Understanding the complicated behavior of quantum many-body systems of interacting bosons
has been a major challenge for leading research groups over the last few years. In fact, 
the growing theoretical interest was further enhanced by recent experimental achievements. 
The most fascinating of these was the realization of Bose-Einstein condensates (BEC) of 
ultra-cold atoms in optical lattices (OL). This allows for the vision that the emerging field
of atomtronics (the atomic analog of quantum electronics) will result in the creation of
a new generation of nanoscale devices. A major advantage of BEC based devices, as compared 
to conventional solid-state structures, lies in the extraordinary degree of precision and 
control that is available, regarding not only the confining potential, but also the strength 
of the interaction between the particles, their preparation, and the measurement of the atomic 
cloud. The realization of atom chips \cite{AK98}, ``conveyor belts'' 
\cite{HHHR01}, atom diodes and transistors \cite{MDJZ04,SAZ07} is considered a major 
breakthrough with potential applications in the field of quantum information processing 
\cite{SFC02}, atom interferometry \cite{SHAWGBSK05} and lasers 
\cite{AK98,MAKDTK97}.


Recently, BECs in driven optical lattices have received a lot of attention. 
It was pointed out \cite{SMSKE04} that the study of the energy
absorption rate (EAR) can be used to probe the quantum phase of the BEC. Specifically,
the excitation spectrum of the system was determined by measuring the EAR induced by a 
periodic modulation of the lattice height. The position and height of the peaks of the 
EAR versus the driving frequency give valuable information about the interaction strength 
and incommensurability of the system. The experimental activity on EAR spectroscopy and 
its applications triggered a theoretical interest in understanding and predicting the EAR
peaks \cite{KIGHS06}. Simultaneously, big efforts were dedicated to the study of
driven dynamics of smaller optical lattices like single and double site systems \cite{SAZ07,JM06,niu,raizen} aiming to understand how to tame quantum dynamics. 

\subsection{Dimers and Trimers}

The theoretical and experimental study of driven dynamics 
in a {\em few site system} using optical lattice technology \cite{AOC05} 
is state of the art. 
So far mainly single and double site (dimer) systems were in the focus 
of actual research. The study of a three-site (trimer)
system adds an exciting topological aspect which we would like 
to explore in this work: the possibility to generate in a controlled 
way circulating atomic currents. 

The possibility to induce DC currents by periodic (AC) modulation of a potential is familiar 
from the context of electronic devices. If an open geometry is concerned, it is referred to 
as ``quantum pumping'' \cite{pumping}, while for closed geometries 
we use the term ``quantum stirring'' \cite{stirring}. In the present paper we consider stirring 
of condensed particles \cite{HKC08a} in a few-site system which is described by the Bose-Hubbard Hamiltonian (BHH).

\subsection{Main observation}

The controlled stirring operation produces an adiabatic DC current in response to a cyclic 
change of control parameters of the optical potential that confines the atoms. We find that 
the nature of the transport process depends crucially on the sign and on the strength of the 
interatomic interactions. We can distinguish four regimes of dynamical behavior. For strong
repulsive interaction the particles are transported one-by-one, which we call \emph{sequential 
crossing}. For weaker repulsive interaction we observe either \emph{gradual crossing} or 
\emph{coherent mega crossing}. Finally, for strong attractive interaction the particles are 
glued together and behave like a huge classical ball that rolls from trap to trap. 

\subsection{Scope and outline}

The present paper has several objectives: {\bf (i)} To introduce an illuminating picture for 
the analysis of transport in a few-site system based on the adiabatic formalism. {\bf (ii)} 
Using this picture to argue that there are four distinct dynamical regimes dependent on the 
strength of the interactions. {\bf (iii)} To propose a controlled stirring process that can 
be utilized in order to probe the strength of the interactions in a trimer system.

The paper is organized as follows: The quantum trimer is introduced in Section~2. We then 
describe qualitatively the stirring process (Section~3), and briefly review the Kubo formula 
approach to quantum pumping~\cite{pmc} (Section~4) which is based on the theory of adiabatic 
processes~\cite{Berry} (see Appendices~A,B). In Section~5 we describe a reduction 
of the three-site Hamiltonian that allows us to regard the stirring of {\em one} particle in 
a trimer as a Landau-Zener (LZ) crossing in a two-level system. The analysis 
is extended to many particles in Section~6 where the variation of the adiabatic energy levels
as a function of the control parameters is analyzed. We regard the transport as a sequence of 
LZ crossings that can or cannot be resolved depending on the strength of the interaction. This 
leads naturally to the distinction between the various regimes of interaction strength. The 
actual calculation of the transport is carried out first for a single particle LZ crossing in 
Sections~7, and in Appendix~C. This is used as a building block for the calculation of the 
stirring in Sections~8,9. The conclusions and perspectives are summarized in Section~10.

\begin{figure}
  \includegraphics[%
  width=0.95\hsize,%
  keepaspectratio]{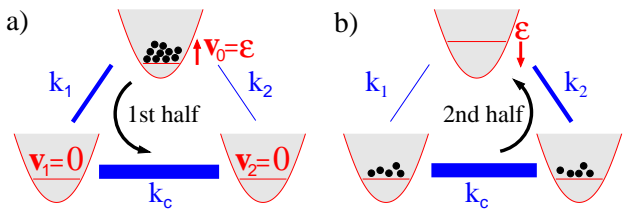}

  \caption{\label{fig:PMB-system}(Color online) Illustration of the model system. Initially,
    all particles are located on the upper site ($i=0$) which represents
    the {}``shuttle''. In the first half of the cycle (a) the on-site
    potential $v_{0}=\varepsilon$ is raised adiabatically slow from a
    very negative initial value and the particles are mainly transported via
    the $k_{1}$ bond to the ``canal'' which is represented by the strongly
    coupled canal sites ($i=1,2$). In the second half of the cycle (b)
    the bias in the coupling is inverted and the particles are mainly transported
    back from the canal to the shuttle via the $k_{2}$ bond. }
\end{figure}


\section{The Bose-Hubbard trimer model}

The simplest model that captures the physics of quantum stirring is the three-site Bose-Hubbard 
Hamiltonian (BHH) \cite{BHK07,HKG06,FP03} (see Fig.~1). This minimal model contains all 
the generic ingredients of large BHH lattices and therefore often is used as a prototype model 
in many recent studies \cite{HKG06,BHK07,SAZ07}. A classical analysis of this model has been performed 
in \cite{FP03,HKG06} where it was shown that for appropriate system parameters and initial 
conditions chaotic dynamics would emerge. In this work we consider adiabatic driving of the 
ground state preparation and therefore chaotic motion is not an issue.

We consider the site index of the quantum trimer taking values $i=0,1,2$. The ${i{=}0}$ site 
has a potential energy~${v_0=\varepsilon}$ and is regarded as a ``shuttle", while the ${i{=}1
{,}2}$ sites are regarded as a two level ``canal" (with ${v_1{=}v_2{=}0}$). The corresponding 
$N$-boson BHH is: 
\be{1}
\mathcal{H}&=&\sum_{i=0}^2 v_i {\hat n}_i+ \frac{U}{2}\sum_{i=0}^{2}{\hat n}_i ({\hat n}_i-1) -
k_c({b}_{1}^{\dagger}{b}_{2}+{b}_{2}^{\dagger}{b}_{1})\nonumber\\ 
&& -k_1({b}_{0}^{\dagger}{b}_{1}+{b}_{1}^{\dagger}{b}_{0}) - 
k_2({b}_{0}^{\dagger}{b}_{2}+{b}_{2}^{\dagger}{b}_{0}).
\label{QMBHH3}
\ee
Without loss of generality we use mass units such that ${\hbar{=}1}$,  
and time units such that intra canal hopping amplitude is ${k_c{=}1}$. 
Accordingly the two single particle levels of the canal 
are $\varepsilon_{\pm}=\pm 1$. 
The annihilation and creation operators ${b}_i$ and ${b}_i^{\dagger}$ obey 
the canonical commutation relations $[{b}_i,{b}_j^{\dagger}]=\delta_{i,j}$ 
while the operators ${\hat n}_i={b}_i^{\dagger}{b}_i$ 
count the number of bosons at site~${i}$. 
The interaction strength between two atoms in a single site is given 
by $U=4 \pi\hbar^2a_sV_{\tbox{eff}}/\mass$ where $V_{\tbox{eff}}$ is the effective volume, 
$\mass$~is the atomic mass, and $a_s$ is the $s$-wave scattering length.

The couplings between the shuttle and the two ends of the canal are $k_1$ and $k_2$. We 
assume that both are much smaller than $k_c$ 
Their inverse $1/k_1$ and $1/k_2$ are like barrier heights, 
and changing them is like switching valves on and off.  
It is convenient to define the two control parameters of the pumping as 
\be{2}
X_1 = \left(\frac{1}{k_2}-\frac{1}{k_1}\right), \qquad
X_2 = \varepsilon.
\ee
By periodic cycling of the parameters $(X_{1},X_{2})$ we can 
induce a circulating current in the system.
We further discuss this controlled stirring process 
in the next section.

\begin{figure}
  \includegraphics[%
  width=\hsize,%
  keepaspectratio]{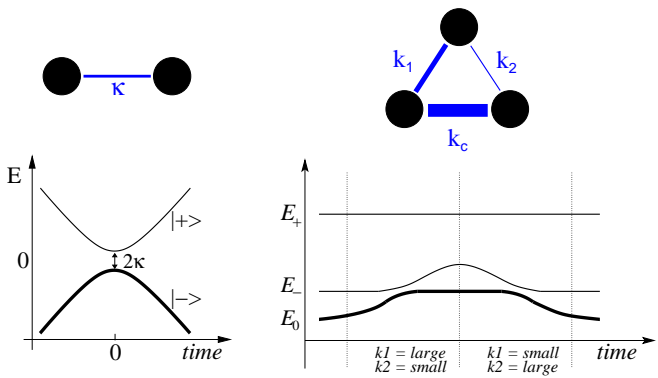} 

  \caption{\label{app2-fig:LZ-crossing} (Color online)
    Scheme of the avoided crossings for the one particle problem. 
    {\em Left panels:} The
    two-site system (\ref{e71}) is prepared
    with one particle in the ground state which initially 
    corresponds to having the particle occupying 
    the left site. As the potential $\varepsilon$
    is raised the particle encounters an avoided crossing. 
    Due to the slowness of the driving it stays in 
    the ground state which implies an adiabatic passage  
    to the right site.
    {\em Right panels:} The three-site trimer system 
    is prepared with one particle in the ground state 
    which initially corresponds to having the particle occupying 
    the shuttle site. As the potential $\varepsilon$
    is varied the particle encounters avoided crossings  
    with the lower canal orbital. A full stirring cycle 
    consists of an adiabatic passage through $k_1$ in the 
    first half of the cycle, and another adiabatic passage 
    through $k_2$ in the second half of the cycle.  
    See the text for further explanations. 
    }
\end{figure}


\section{Stirring}

By periodic cycling of the parameters $(X_1,X_2)$ we can imitate a classical peristaltic 
mechanism and obtain a non-zero amount ($Q$) of transported atoms per cycle. During the 
driving cycle the total number of bosons remains constant. The energy is not a constant 
of motion, but in the adiabatic limit considered here, the system returns to the same 
state at the end of each cycle.

The pumping cycle is illustrated in Figs.~1-3. Initially all the particles are located in 
the shuttle which has a sufficiently negative on-site potential energy ($X_2<0$). In the 
first half of the cycle the coupling is biased in favor of the $k_1$ route (${X_1>0}$) while 
$X_2$ is raised until (say) the shuttle is empty. In the second half of the cycle the coupling 
is biased in favor of the $k_2$ route (${X_1<0}$), while $X_2$ is lowered until the shuttle 
is full. Assuming~$U{=}0$, the shuttle is depopulated via the $k_1$ route into the lower 
energy level $\varepsilon_{-}$ during the first half of the cycle, and re-populated via the 
$k_2$ route during the second half of the cycle. Accordingly the net effect is to have a 
non-zero~$Q$. 

If we had a single particle in the system, the net effect would be to pump roughly one 
particle per cycle. If we have~$N$ non-interacting particles, the result of the same cycle 
is to pump roughly~$N$ particles per cycle. 
We would like to know what is the actual result using 
a proper quantum mechanical calculation, 
and furthermore we would like to investigate what is the 
effect of the interatomic interaction~$U$ on the result.

\begin{figure}
  \includegraphics[%
  width=0.8\hsize,keepaspectratio]{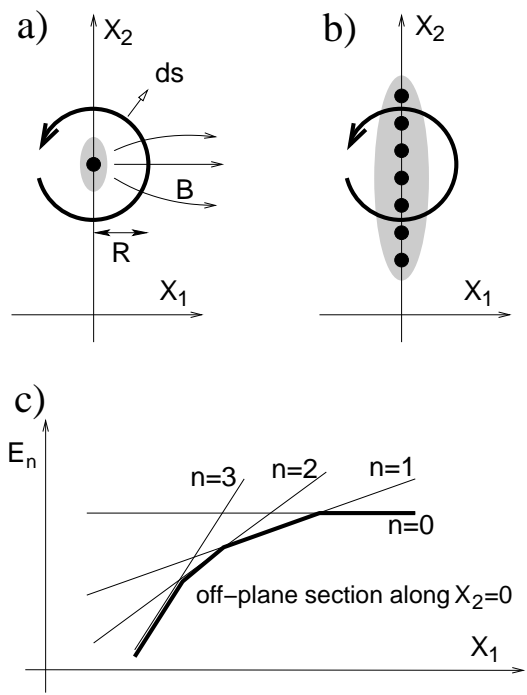}

  \caption{\label{fig:PMB-cycle}
    Analysis of the pumping cycle for $N$~particles. See
    the text for further details. For a large cycle that encircles the
    whole shaded region we have $Q\approx N$. The position of the ``monopoles''
    is depicted by black dots: (a) no interactions (all monopoles are
    ``piled up'' at the same position) (b) with interactions. In panel (c) we schematically
    plot the energy levels along the $X_{1}=0$ axis for a system corresponding
    to $N=3$ bosons. Note that energy levels that correspond to 
    non-participating states (those with non-zero occupation of 
    the upper canal orbital) are not plotted.}
\end{figure}

The above description of the stirring process might look  
convincing, but in fact it does not hold in the quantum 
mechanical reality. The quantum stirring process 
is in general not a peristaltic process but rather 
a coherent transport effect. This point is best clarified 
by observing that the simple minded picture above 
implies that the amount of pumped particles per cycle 
is at most~$N$. This conclusion is wrong. We shall explain 
in the next section that in principle one can get ${Q \gg N}$ 
per cycle. The proper way to think about the quantum stirring 
process is as follows: Changing a control parameter, say $X_2$ 
with some constant rate $\dot{X}_2$, induces a circulating 
current in the system. Each particle can encircle the system 
more than once during a full cycle. Hence ${Q \gg N}$ is feasible.


\section{The adiabatic picture}

In analogy with Ohm's law (where $X$ is the magnetic flux, and $-\dot{X}$ is the electro motive 
force), the current is  ${I=-G_1\dot{X}_1}$ if we change~$X_1$ and  ${I=-G_2\dot{X}_2}$ if we 
change~$X_2$, where $G_1$ and $G_2$ are elements of the geometric conductance matrix. Accordingly       
\be{3}
Q = \oint\limits_{\tbox{cycle}} Idt 
= -\oint (G_1 dX_1 + G_2 dX_2).
\ee     
In order to calculate the geometric conductance we use the Kubo formula approach to quantum pumping 
\cite{pmc} which is based on the theory of adiabatic processes \cite{Berry}. It turns 
out that in the strict adiabatic limit $G$ is related to the vector field $\mathbf{B}$ also known as 
``two-form" in the theory of Berry phase. Namely, using the notations ${\mathbf{B}_1=-G_2}$ and 
${\mathbf{B}_2=G_1}$ we can rewrite Eq.(\ref{e3}) as 
\be{7}
Q = \oint \mathbf{B} \cdot d\vec{s} ,
\ee     
where we define the normal vector $d\vec{s}=(dX_2,-dX_1)$ as illustrated in Fig.~3. The advantage 
of this point of view is in the intuition that it gives for the result: $Q$ is related to the flux 
of a field~$\mathbf{B}$ which is created by ``magnetic charges" in $X$~space. For $U{=}0$ all the 
magnetic charge is concentrated in one point. As the interaction~$U$ becomes larger the ``magnetic 
charge" disintegrates into $N$ elementary ``monopoles" (see Fig.~3). In practice the calculation 
of $\mathbf{B}$ is done using the following formula:   
\be{8}
\mathbf{B}_j=\sum_{n\ne n_0}
\frac{2 \ {\rm Im}[\mathcal{I}_{n_0n}] \ \mathcal{F}^j_{nn_0}}
{(E_n-E_{n_0})^2},
\ee
where the current operator is conveniently defined as  
\be{0}
\mathcal{\mathcal{I}}
&\equiv&  \frac{1}{2}(\mcal{I}_{0\mapsto1}+\mcal{I}_{2\mapsto0}) \nonumber \\
&=&
\frac{i}{2}
\left[
k_1({b}_{0}^{\dagger}{b}_{1}-{b}_{1}^{\dagger}{b}_{0})
+k_2({b}_{2}^{\dagger}{b}_{0}-{b}_{0}^{\dagger}{b}_{2})
\right],\label{eq:I-operator}\ee
while the generalized force operators are defined as
\be{0}
\mathcal{F}^j \ \ = \ \ -\frac{\partial \mathcal{H}}{\partial X_j}
\ee 
and are associated with the control parameters $X_j$. The index $n$ distinguishes the eigenstates 
of the many-body Hamiltonian. We assume from now on that $n_0$ is the BEC ground state. 


\begin{figure*}
\includegraphics[%
  width=\hsize,
  keepaspectratio]{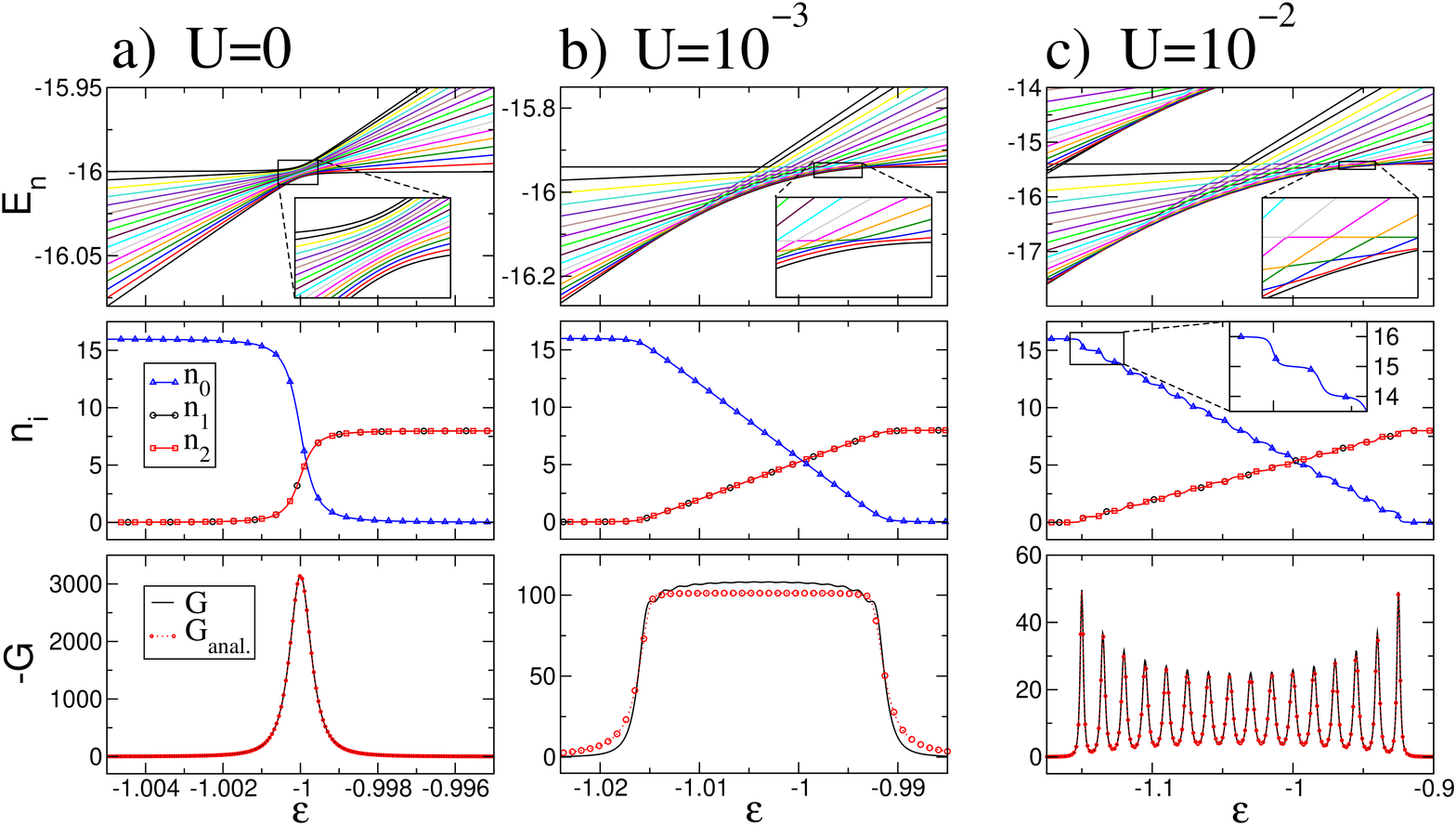}

\caption{\label{fig:E-n-G1} (Color online) Evolution of the energy levels, the site
occupation and the conductance for $N=16$ particles and
$k_1=2\times10^{-4},\;k_2=1\times10^{-4}$. We refer to three representative
values of $U$, which are indicated on top of each set of panels.
Upper panels: the lowest $N+1$ energy levels $E_{n}$ which dominate
the conductance $G_{2}$ are plotted as a function of $X_{2}=\varepsilon$.
The insets represent magnifications of the indicated areas. Middle
panels: the site occupations $n_{0}$(blue $\Delta$), $n_{1}$(black
$\circ$), $n_{2}$(red $\Box$). Note the steps of size $1$ for
the dot-occupation and $1/2$ in the wire-sites occupation in subfigure
(c). Lower panels: the corresponding conductance $G_{2}$ as a function
of $\varepsilon$. Numerical results are represented by solid black
lines while the dotted red line corresponds to the analytical result
(\ref{eq:G-anal-mega}) in (a) and to (\ref{eq:G-anal-sequential})
in (b), (c).}

\end{figure*}

\section{The two-orbital approximation}

We assume an adiabatic process that involves only two orbitals: 
the shuttle orbital and the orbital of the lower canal level (see
right panel of Fig.~2). 
Accordingly we can simplify the Hamiltonian in a way that 
illuminates the physics of the quantum stirring process
and simplifies the formal treatment. For pedagogical reasons
we consider first the single particle ($N{=}1$) case, and
extend the calculation to the many-body case in the next
section. The model Hamiltonian in the position basis is: 
\be{0}
  \mathcal{H}=\left(\begin{array}{ccc}
      \varepsilon & -k_{1} & -k_{2}\\
      -k_{1} & 0 & -1\\
      -k_{2} & -1 & 0\end{array}\right).
\ee
The current can be measured on the $0\mapsto1$ bond 
or on the $2\mapsto0$ bond, accordingly:
\be{0}
  \mathcal{\mathcal{I}}_{0\mapsto1} &=& \left(\begin{array}{ccc}
      0 & -ik_{1} & 0\\
      ik_{1} & 0 & 0\\
      0 & 0 & 0\end{array}\right),
\\
  \mathcal{\mathcal{I}}_{2\mapsto0} &=& \left(\begin{array}{ccc}
      0 & 0 & ik_{2}\\
      0 & 0 & 0\\
      -ik_{2} & 0 & 0\end{array}\right).
\ee
For zero couplings (${k_{1}=k_{2}=0}$) the 
eigenenergies of the orbitals 
are ${\varepsilon_{\pm}=\pm 1}$ 
and ${\varepsilon_{0} = \varepsilon}$. 
The corresponding eigenstates are 
\be{0}
|\varepsilon_{0}\rangle=\left(\begin{array}{c}
1\\
0\\
0\end{array}\right)
,\,\,\,\,\,\,
|\varepsilon_{-}\rangle=\frac{1}{\sqrt{2}}\left(\begin{array}{c}
0\\
1\\
1\end{array}\right)
,\\
|\varepsilon_{+}\rangle=\frac{1}{\sqrt{2}}\left(\begin{array}{c}
0\\
1\\
-1\end{array}\right).
\ee
The Hamiltonian and the current operators in this orbital basis are: 
\be{0}
\mathcal{H}=\left(\begin{array}{ccc}
\varepsilon & -\kappa  & -\lambda\kappa\\
-\kappa  & -1 & 0\\
-\lambda\kappa & 0 & 1\end{array}\right)
\ee
and
\be{0}
\mathcal{\mathcal{I}}_{0\mapsto1}
=\frac{k_{1}}{k_{1}+k_{2}} 
\left(\begin{array}{ccc}
0 & -i\kappa & -i\kappa\\
i\kappa & 0 & 0\\
i\kappa & 0 & 0\end{array}\right),
\\
\mathcal{\mathcal{I}}_{2\mapsto0}
=\frac{k_{2}}{k_{1}+k_{2}}
\left(\begin{array}{ccc}
0 & i\kappa & -i\kappa\\
-i\kappa & 0 & 0\\
i\kappa & 0 & 0\end{array}\right).
\ee
where the effective coupling between 
the shuttle orbital and the lower canal 
orbital is 
\be{0}
\kappa = \frac{k_{1}+k_{2}}{\sqrt{2}}
\ee
and the net ``splitting ratio" is \cite{cnb}  
\be{33}
\lambda   = \frac{k_{1}-k_{2}}{k_{1}+k_{2}}.
\ee
The stirring cycle starts with all particles localized 
in the shuttle orbital and the adiabatic particle transport
takes place during the avoided crossing of this orbital 
with the lower canal orbital. 
Accordingly, we focus on the upper left ($2\times2$) submatrix:
\be{0}
\mathcal{H}=\left(\begin{array}{cc}
\varepsilon & -\kappa \\
-\kappa  & -1\end{array}\right).
\ee
In practice it is more convenient to define the averaged current 
operator ${\mcal{I}\equiv(\mcal{I}_{0\mapsto1}+\mcal{I}_{2\mapsto0})/2}$ 
whose matrix representation is 
\be{0}
\mcal{I} 
=
\frac{\lambda}{2}
\left(\begin{array}{cc}
0 & -i\kappa\\
i\kappa & 0\end{array}\right).
\ee
The advantage of this definition is that within the two halves of
a symmetric pumping cycle, while $\varepsilon$ is raised or lowered,  
the same amount of particles is being transported. 
Thus in order to get $Q$ for a full cycle,  
we simply double the integrated current over a half cycle, 
where the variation of the control parameter $\varepsilon$ is monotonic 
starting from a very negative initial value. 
For completeness we also write the matrix 
representation of the generalized force that is 
conjugate to ${X_2=\varepsilon}$  
\be{0}
\mathcal{F}
=
\left(\begin{array}{cc}
1 & 0\\
0 & 0\end{array}\right).
\ee
The above expressions for $\mathcal{H}$ and $\mathcal{I}$ and $\mathcal{F}$ 
completely define the transport problem in the case of one particle. 
Within the two orbital approximation the calculation of~$Q$ therefore
reduces to the study of a single LZ crossing, which we discuss in
Section~VII.


\section{The adiabatic variation of the many-body levels}

We now turn to the analysis of the many-body problem. 
The first step is to understand the evolution
of the eigenenergies $E_{n}$ as $\varepsilon$ is varied, 
while the other parameters are kept constant. 
It is convenient to rewrite the BHH as 
\be{0}
\mathcal{H} & = & \mathcal{H}_{\tbox{shuttle}}(\varepsilon)+\mathcal{H}_{\tbox{canal}}+\mathcal{H}_{\tbox{cpl}}\label{eq:BHH-shuttle-canal}
\\
\mathcal{H}_{\tbox{shuttle}} 
& = & 
\frac{U}{2}\hat{n}_{0}(\hat{n_0}{-}1)+\varepsilon\hat{n}_{0}
\nonumber
\\
\mathcal{H}_{\tbox{canal}} 
& = & 
\frac{U}{2}[\hat{n}_{1}(\hat{n}_{1}{-}1)+\hat{n}_{2}(\hat{n}_{2}{-}1)]-(\dog{{b}_{2}}{b}_{1}+\dog{{b}_{1}}{b}_{2})
\nonumber
\\
\mathcal{H}_{\tbox{cpl}} 
& = & 
-k_{1}(\dog{{b}_{0}}{b}_{1}+\dog{{b}_{1}}{b}_{0})-k_{2}(\dog{{b}_{0}}{b}_{2}+\dog{{b}_{2}}{b}_{0}),\nonumber
\ee
while the operators for the current and the generalized force are
\be{0}
\mathcal{\mathcal{I}}_{0\mapsto1}&=&ik_{1}(\dog{{b}_{1}}{b}_{0}-\dog{{b}_{0}}{b}_{1})\nonumber\\
\mathcal{\mathcal{I}}_{2\mapsto0}&=&ik_{2}(\dog{{b}_{0}}{b}_{2}-\dog{{b}_{2}}{b}_{0})\nonumber\\
\mathcal{\hat{F}}&=&-\hat{n}_{0}.
\ee
In what follows we assume $0<k_{1},k_{2}\ll k_{c}=1$, 
and $N|U|\ll k_{c}=1$, and consequently generalize 
the ``two orbital approximation'' of the previous section 
to the case of ${N>1}$ particles.

\begin{figure*}
  \includegraphics[%
  width=0.65\hsize,
  keepaspectratio]{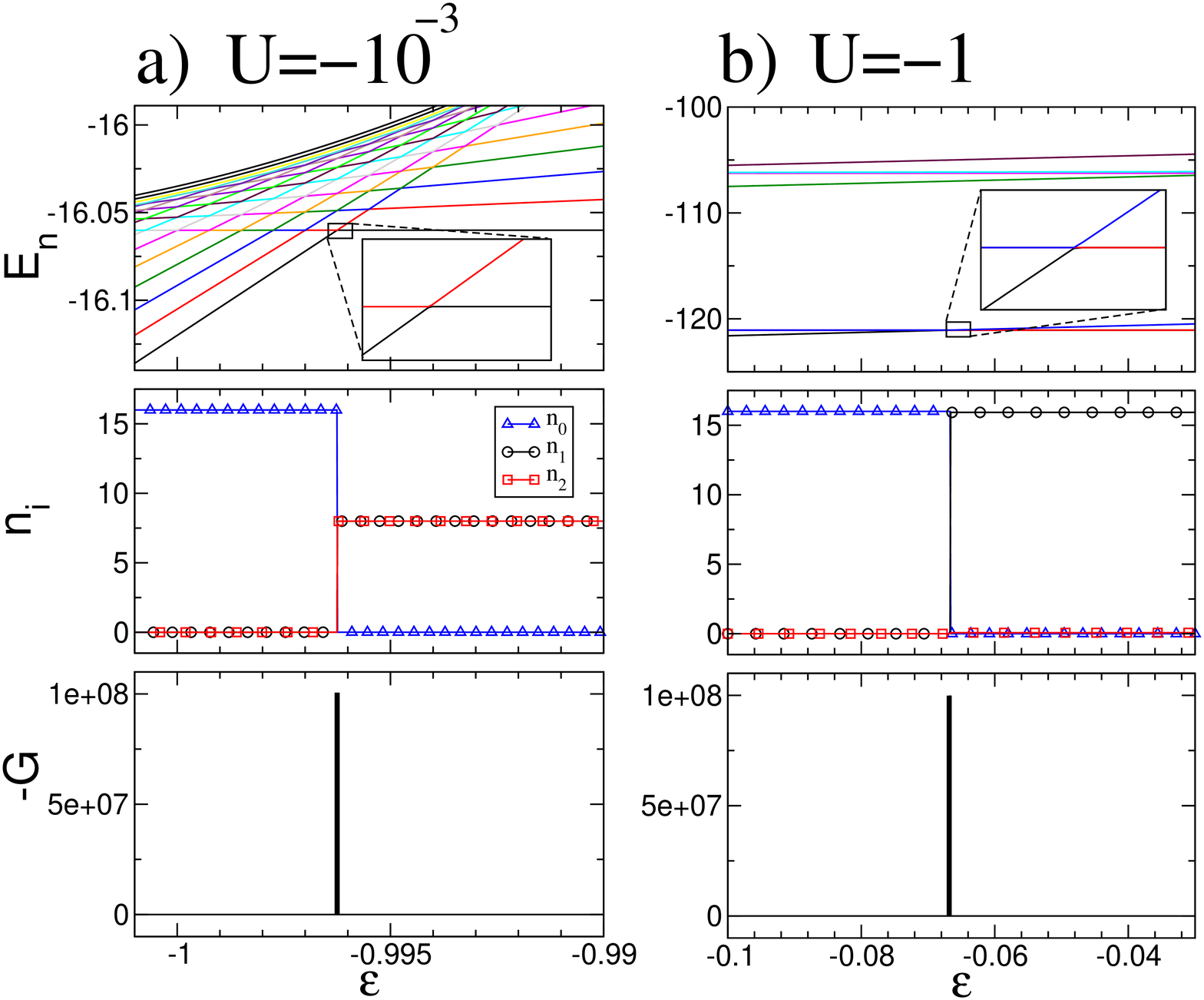}

  \caption{\label{fig:E-n-G2}(Color online) Evolution of the energy levels, the site
    occupation and the conductance for strong attractive interaction $U=-1$
    (see Fig.~\ref{fig:E-n-G1} for legend and parameters). The particles
    are ``glued    together'' and roll like a classical ball from the shuttle
    to the left wire site as can be seen from the middle panel where the site
    occupations $n_{0}$(blue $\Delta$), $n_{1}$(black $\circ$), $n_{2}$(red
    $\Box$) are plotted. The width of the transition is exponentially  
    small in~$N$, as explained in the text, and therefore cannot be resolved numerically.}
\end{figure*}

In the zeroth order approximation $k_1$ and $k_2$ are neglected;  
later we take them into account as a perturbation. 
For ${k_1=k_2=0}$ the number ($n$) of particles 
in the shuttle becomes a good quantum
number.  The other ${N{-}n}$ particles occupy 
the lower orbital ($\varepsilon_{-}$)  
of the canal because we assume $NU \ll k_c$.
Hence the many-body energies are 
\be{0}
E_n = E_{\tbox{shuttle}}(n) + E_{\tbox{canal}}(N{-}n),
\ee
where $n{=}0,1,..,N$, and 
\be{0}
E_{\tbox{shuttle}} &=& \varepsilon n + \frac{1}{2}U (n{-}1)n \\
E_{\tbox{canal}} &=& -(N{-}n) + \frac{1}{4}U(N{-}n{-}1)(N{-}n).
\ee
From the degeneracy condition ${E_{n}-E_{n{-}1}=0}$ 
(${n{=}1,2,...N}$ is the number of particles in the
shuttle) we determine the location of the 
${n\mapsto (n{-}1)}$ crossing to be
\be{0}
\varepsilon_n=-1 + \frac{1}{2}U \times (N{-}3n{+}2).
\ee
Accordingly, we conclude that the $N$ crossings are distributed within
\be{11}
-1-(N{-}1)U \ \ \le \ \ \varepsilon \ \ \le \ \ -1+\frac{1}{2}(N{-}1)U.
\ee
One can introduce a rescaled control variable $\hat{\varepsilon}$ which reads
\be{12}
\hat{\varepsilon} = \frac{\varepsilon+1}{(N-1)U},
\label{eq:epsilon-rescaled}
\ee
and its support is ${-1 < \hat{\varepsilon} < 1/2}$. The distance between the crossings, while 
varying the shuttle potential $\varepsilon$, is $(3/2)U$. Once we take $\kappa$ into account we 
get {\em avoided} crossings, whose width we will estimate in the next paragraph.

Within the framework of the two-orbital approximation, the truncated many-body Hamiltonian matrix 
takes the form  
\be{0}
\mathcal{H}_{nm}= E_n\delta_{n,m} - \kappa_n\delta_{n,n\pm1},
\ee
where $n{=}0,\cdots,N$ and the couplings are defined 
as ${\kappa_n=\langle n{-}1|\mathcal{H}|n\rangle}$. 
For example, in the ${N{=}3}$ case we have
\be{0}
\mathcal{H}=\left(
\begin{matrix}
E_{0} & -\kappa _{1} & 0 & 0 \\
-\kappa _{1} & E_{1} & -\kappa _{2} & 0 \\
0 & -\kappa _{2} & E_{2} & -\kappa _{3} \\
0 & 0 & -\kappa _{3} & E_{3}  
\end{matrix}
\right).
\ee
\hide{
\be{0}
\mathcal{H}=\left(
\begin{matrix}
E_{0} & -\kappa _{1} & 0 & 0 & 0\\
-\kappa _{1} & E_{1} & -\kappa _{2} & 0 & 0\\
0 & -\kappa _{2} & E_{2} & -\kappa _{3} & 0\\
0 & 0 & -\kappa _{3} & E_{3} & -\kappa _{4}\\
0 & 0 & 0 & -\kappa _{4} & E_{4}
\end{matrix}
\right)
\ee
}
The calculation of ${\kappa_n}$ involves the matrix elements of $b_i^{\dag}b_0$, leading to 
\be{0}
\kappa_n = [(N+1-n)n]^{1/2} \,\kappa.
\ee 
An analogous expression applies to the truncated current operator:
\be{0}
\mathcal{I}=
\frac{1}{2}
\left(
\begin{matrix}
0 & -i\lambda_1\kappa _{1} & 0 & 0 \\
i\lambda_1\kappa _{1} & 0 & -i\lambda_2\kappa _{2} & 0 \\
0 & i\lambda_2\kappa _{2} & 0 & -i\lambda_3\kappa _{3} \\
0 & 0 & i\lambda_3\kappa _{3} & 0 
\end{matrix}
\right).
\ee
\hide{
\be{0}
\mathcal{I}=
\frac{1}{2}
\left(
\begin{matrix}
0 & -i\lambda_1\kappa _{1} & 0 & 0 & 0\\
i\lambda_1\kappa _{1} & 0 & -i\lambda_2\kappa _{2} & 0 & 0\\
0 & i\lambda_2\kappa _{2} & 0 & -i\lambda_3\kappa _{3} & 0\\
0 & 0 & i\lambda_3\kappa _{3} & 0 & -i\lambda_4\kappa _{4}\\
0 & 0 & 0 & i\lambda_4\kappa _{4} & 0
\end{matrix}
\right)
\ee
}

\begin{figure}
\includegraphics[%
  width=\hsize,
  keepaspectratio]{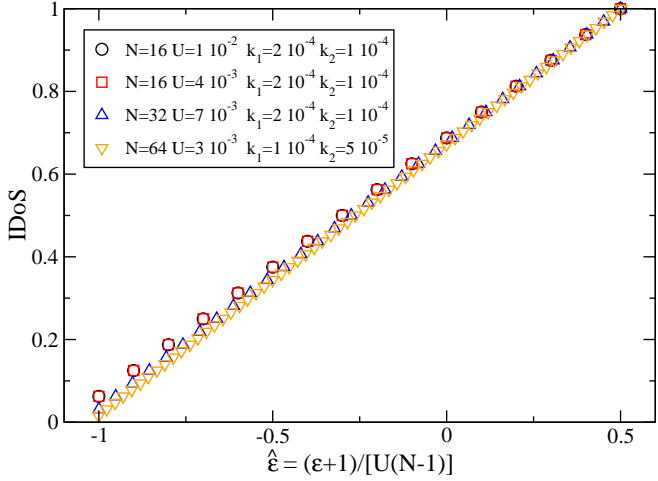}

  \caption{\label{fig:IDOS-scaling}(Color online) Integrated density (IDoS) of avoided
    crossings ({}``magnetic monopoles'') for various values of the parameters
    $U$, $k_{1}$, $k_{2}$ and the boson number $N$ as a function of
    the rescaled on-site potential $\hat{\varepsilon}$. The support of
    the IDoS is predicted by Eq.~(\ref{eq:epsilon-rescaled}) to
    be $\hat{\varepsilon}=[-1,0.5]$ which is nicely confirmed.}
\end{figure}

\begin{figure}
\includegraphics[%
  width=\hsize,
  keepaspectratio]{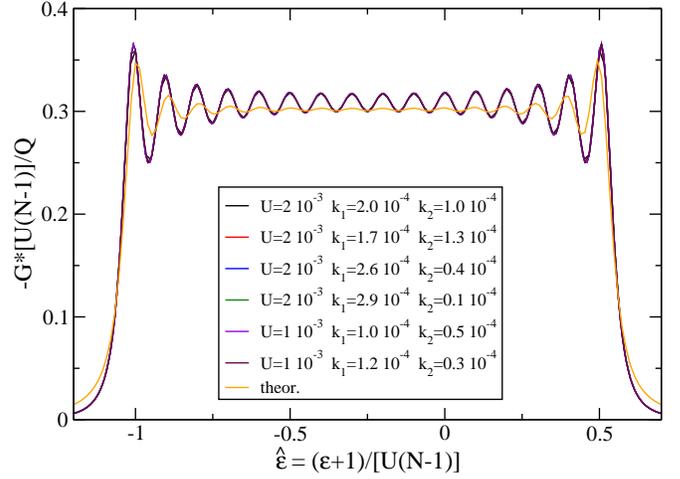}

\caption{\label{fig:G-scaled}
(Color online) Scaling behavior of the conductance $G$ 
in the gradual crossing regime ($U/\kappa\approx4.7$). 
We have $N=16$ particles, and the curves correspond 
to various values of $\kappa$ and $\lambda$.    
The $x$-axis is the rescaled control variable $\hat{\varepsilon}$ (\ref{eq:epsilon-rescaled})
and the $y$-axis is scaled accordingly to preserve the net charge.
Additionally, the ordinate was scaled by the expected charge $Q$
for a half-cycle according to Eq.~(\ref{eq:Q-anal}) and indeed
the area under the curve is $Q\approx0.5$. One observes that the
curves fall on top of one another which confirms the dependence of
$G$ on the ratio $U/\kappa$. Additionally, we overplotted the theoretical
expression (solid orange line) for the conductance (\ref{eq:G-anal-sequential}).
Although this expression is not expected to be valid in the intermediate
regime, the agreement is pretty good. Also the agreement with the
estimation from Eq.~(\ref{eq:G-anal-intermediate}) corresponding
to a constant value of $-G\approx0.315$ is apparent.}
\end{figure}

For large $U$, as $\varepsilon$ is varied, we encounter a sequence of distinct LZ transitions:  
\be{0}
|3\rangle\,\,\,\,\overset{\kappa _{3}}{\longmapsto}\,\,\,\,|2\rangle\,\,\,\,\overset{\kappa _{2}}{\longmapsto}\,\,\,\,|1\rangle\,\,\,\,\overset{\kappa _{1}}{\longmapsto}\,\,\,\,|0\rangle
\ee
The distance between avoided crossings is of order $U$ while their width is 
\be{0}
\delta \varepsilon_n = \kappa_n.
\ee
The widest crossings are at the center with ${\delta \varepsilon_n \sim N\kappa}$. This width should
be smaller than the spacing~$U$ between avoided crossings, else they merge and we no longer
have distinct crossings. The other extreme possibility is to regard $U$ as the perturbation
rather than $\kappa$. The width of the one-particle crossing is $\kappa$, and it would not be
affected by the many-body interaction as long as the span $NU$ is much smaller than that.  
We therefore deduce that for repulsive interaction there are three distinct regimes: 
\be{14}
U \ll \kappa/N & \ \ \ \ & \mbox{mega crossing regime} \nonumber\\ 
\kappa/N < U < N\kappa & \,\,\,\,\,\,\,\, & \mbox{gradual crossing regime} \\
U \gg N\kappa & \ \ \ \ & \mbox{sequential crossing regime}\nonumber
\ee 
Accordingly, depending on the ratio $U/\kappa$ 
we expect different results for $G_2(X)$. 
Indeed in a later section this expectation is confirmed 
both analytically and numerically (Figs.~\ref{fig:E-n-G1}-\ref{fig:E-n-G2}).
In Fig.~\ref{fig:IDOS-scaling} we report the integrated density 
of avoided crossings (IDoS) for various values of $U$, $\kappa$, $\lambda$ 
and number of bosons~$N$. We find that all points fall in the 
predicted range confirming nicely the scaling relation (\ref{eq:epsilon-rescaled}). 
If $U/\kappa$ is small these avoided crossings merge and cannot be resolved.   
In a later section we discuss the implied scaling relation 
for the conductance (Fig.~\ref{fig:G-scaled}). 

\section{Transport during a LZ crossing}

The prototype example for an adiabatic crossing is the LZ problem. In this section we discuss 
the analysis of the transport during a LZ crossing, while in the next section we shall use the 
obtained result as a building block for the analysis of the transport during a stirring process. 
The following  treatment assumes a strict adiabatic process. A more advanced treatment that takes 
into account non-adiabatic transitions can be found in \cite{cnb}, where also the resulting
fluctuations in~$Q$ are calculated. 

Consider a single particle in a two-site system (see left panels of Fig.~2). The coupling between the sites is $\kappa$ 
while the on-site potentials are $v_{i}=\pm \varepsilon/2$, where ${i=1,2}$ labels the sites. 
The Hamiltonian is  
\be{71}
\mathcal{H}_{ij}=\left(\begin{array}{cc}
\varepsilon/2 & -\kappa \\
-\kappa  & -\varepsilon/2\end{array}\right).
\ee
At time $t=-\infty$, the control parameter ${X=\varepsilon}$ is very negative
and the particle is on the left site ($i=1$). Then~$X$
is increased adiabatically and the levels experience an avoided crossing
leading to the adiabatic transfer of the particle 
to the right site ($i=2$). The current operator and the generalized 
force operator are represented by the matrices 
\be{72}
\mcal{I}_{ij} 
=
\left(\begin{array}{cc}
0 & -i\kappa\\
i\kappa & 0\end{array}\right)
\ee
and 
\be{73}
\mathcal{F}_{ij}
=
\left(\begin{array}{cc}
1/2 & 0\\
0 & -1/2\end{array}\right).
\ee
The instantaneous eigenenergies of the LZ Hamiltonian 
are labeled as ${n={-}}$ (lower) and ${n{=}+}$ (upper):  
\be{0}
E_n  =  \mp\frac{1}{2}\Omega
\ee
and the associated eigenstates are
\be{74}
|E_{-}\rangle & = & \left(\begin{array}{c}
\cos({\theta}/{2})\\
\sin({\theta}/{2})\end{array}\right)\nonumber \\
|E_{+}\rangle & = & \left(\begin{array}{c}
-\sin({\theta}/{2})\\
\cos({\theta}/{2})\end{array}\right),
\ee
where 
\be{0}
\Omega & = & \sqrt{\varepsilon^{2}+(2\kappa)^{2}}\\
\theta & = & -\arctan\left({2\kappa}/{\varepsilon}\right).
\ee
Note that~$\theta{=}0$ at ${t{=}-\infty}$ evolves to~$\theta{=}\pi$ at ${t{=}\infty}$. The 
matrix representation of $\mathcal{I}$ and $\mathcal{F}$ in this basis is
\be{0}
\mcal{I}_{nm} 
=
\left(\begin{array}{cc}
0 & -i\kappa\\
i\kappa & 0\end{array}\right),
\ee
and 
\be{0}
\mathcal{F}_{nm}
=
\frac{1}{2}
\left(
\begin{array}{cc}
\cos(\theta) & -\sin(\theta)\\
-\sin(\theta) & -\cos(\theta)
\end{array}\right).
\ee
Now we can use Eq.(\ref{e8}) to obtain 
the geometric conductance:
\be{0}
\label{Ggeom1}
G(\varepsilon)
= \frac{\kappa \sin(\theta) }{\Omega^2}
=-\frac{2\kappa^{2}}{[\varepsilon^{2}+(2\kappa)^{2}]^{3/2}}.
\ee
Since we assume here a strictly adiabatic process 
we expect $100\%$ transfer efficiency, and indeed:
\be{-1}
Q 
\ &=& \   
\int\langle\mathcal{I}\rangle dt
\ = \ -\int Gd\varepsilon
\\ 
\ &=& \ 
\left.
\frac{\varepsilon}{2\sqrt{\varepsilon^{2}+(2\kappa)^{2}}}
\right|_{-\infty}^{\infty}
\ = \ 1.
\ee
For pedagogical reasons, we rederive the result (\ref{Ggeom1}) for the conductance 
$G(\varepsilon)$ in a straightforward way from the Schr\"odinger equation in Appendix C.
This might add to the understanding of the general discussion of quantum stirring
presented in a later section.

\section{Transport during stirring:\,\,\,\,\,\,\, the~case~${\mathbf U=0}$}

We have realized that the stirring problem of one particle 
reduces to the LZ-like crossing problem Eqs.(\ref{e71},\ref{e72}). 
Disregarding the different on-site 
energies, the Hamiltonian is the same as in the LZ-problem Eq.(\ref{e71}). 
The significant difference is related to the current operator. 
Its multiplication by the splitting ratio $\lambda$  
is the fingerprint of the non-trivial topology. 
We further discuss the physics behind $\lambda$ at the end of this section.
On the technical side, all we have to do is to multiply the 
result obtained in the previous section by this 
factor. If we have $N$~non~interacting Bosons, the result 
should be further multiplied by $N$, leading to    
\be{0}
G(X_1,X_2)
=-\frac{1}{2}N\lambda
\frac{2\kappa^{2}}{[(\varepsilon-\varepsilon_{-})^{2}+(2\kappa)^{2}]^{3/2}}.
\label{eq:G-anal-mega}
\ee
In this expression $\varepsilon$ is determined by $X_2$ 
and $\lambda$ is determined by $X_1$, 
while $\kappa$ is conveniently regarded as a fixed parameter. 
For a full cycle we get 
\be{0}
Q
\ \ = \ \ 
\oint\langle\mathcal{I}\rangle dt
\ \ = \ \ 
N\lambda . 
\label{eq:Q-anal}
\ee
We note that this result assumes a symmetric 
stirring cycle such that ${k_1=k_{\tbox{large}}}$ 
and ${k_2=k_{\tbox{small}}}$ in the first half of 
the cycle, while ${k_1=k_{\tbox{small}}}$   
and ${k_2=k_{\tbox{large}}}$ in the second half 
of the cycle. Accordingly
\be{0}
Q
\ \ = \ \ 
N\frac{k_{\tbox{large}}-k_{\tbox{small}}}{k_{\tbox{large}}+k_{\tbox{small}}}.
\ee
It is more illuminating to denote the $X_1$-radius of the pumping cycle by $R$ , as illustrated 
in Fig.~\ref{fig:PMB-cycle} and to re-write the latter expression as follows: 
\be{0}
Q=N\frac{[1+(\kappa R)^{2}]^{1/2}-1}{\kappa R}.
\ee
In particular for small cycles we get a linear dependence on the radius   
\be{81}
Q \ \ \approx \ \ N \kappa R, 
\ee
while for large cycles we obtain the limiting value   
\be{82}
Q \ \ \approx \ \ N.
\ee

In a two-site topology the amount of particles that 
are transported during a strictly adiabatic LZ crossing 
is exactly~$N$. In contrast to that, in the stirring 
problem that we study, we have non trivial ``ring" topology, 
and therefore $Q$ is multiplied by the splitting ratio~$\lambda$.
In practice $k_1$ and $k_2$ are positive and 
accordingly ${|\lambda|<1}$. But in principle 
we can have ${|\lambda|>1}$.  This would happen 
if $k_1$ and $k_2$ were negative. In such case 
the lower orbital of the ``canal"
is anti-symmetric rather than symmetric. 
Going through the derivation one observes that 
the same results apply with 
\be{0}
\lambda \ \ \mapsto \ \ 1/\lambda .
\ee
For small cycles we find
\be{83}
Q \ \ \approx \ \ N \left[\kappa R \right]^{-1},
\ee
which means that we can circulate ${Q\gg N}$ particles per cycle. This demonstrates our statement 
that quantum stirring is not a classical-like peristaltic process, but rather a coherent transport 
effect.

The results above become more transparent if we 
notice that they reflect the topology 
of the $(X_1,X_2,X_3)$ space, 
where $X_3$ is a fictitious Aharonov-Bohm flux  
to which the operator $\mathcal{I}$ is conjugate 
[see Appendix~A].
In this extended space the field $\mathbf{B}$ 
that appears in Eq.(\ref{e7}) has zero divergence,  
with the exception  of the ``Dirac monopoles"   
which are located at points where $E_{n_0}$ 
has a degeneracy with a nearby level. 
If the shuttle orbital and the lower orbital
of the ``canal" have the opposite parity 
this degeneracy point is located 
at ${X=(0,\varepsilon_{-},0)}$ which is 
in the plane of the pumping cycle. 
But if the shuttle orbital and the lower orbital  
of the ``canal" have the same parity, 
then this degeneracy point is displaced off-plane. 
Accordingly we get the divergent result Eq.(\ref{e83}) 
or the non-divergent result Eq.(\ref{e81}).
It is important to realize that because 
of the gauge invariance under the transformation $X_3 \mapsto X_3{+}2\pi$ 
the degeneracy point is duplicated, 
leading to a ``Dirac chain". 
In the far field  this chain looks like 
a charged line, and accordingly in the far 
field we get Eq.(\ref{e82}) which in leading 
order is not sensitive to the parity of the orbitals.

If we had only one particle, the degeneracy at 
the center of the pumping cycle would correspond 
to a single Dirac monopole. If we have $N$ non-interacting
particles we have in fact $N$  Dirac monopoles 
at the same location. As the interaction~$U$ 
is turned on this ``pile" disintegrates 
into $N$ elementary ``monopoles" (see Fig.~3).  
This will be further discussed in the next section.

\begin{figure}
\includegraphics[%
  width=\hsize,
  keepaspectratio]{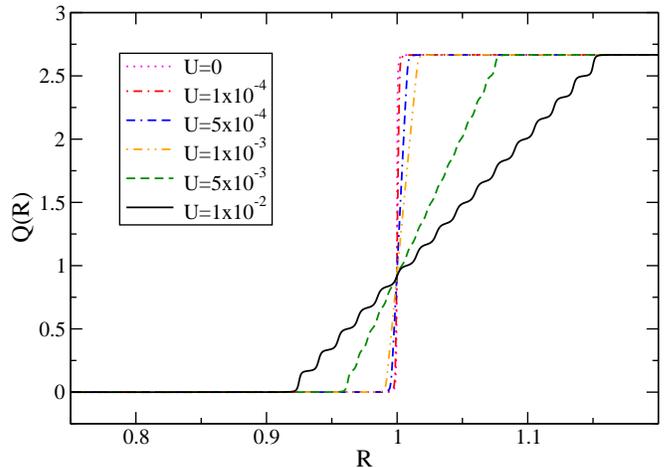}

\caption{\label{fig:QvsR} (Color online) Dependence of~$Q$ on the $X_2$-radius~$R$ 
of the pumping cycle for the system presented in Fig.~\ref{fig:E-n-G1}.
The curves correspond to different values of the interaction strength~$U$
and represent the net amount of particles ($Q$) transported during the
first half of a symmetric pumping cycle centered around $X_2=\varepsilon=0$.
\hide{The lines by order are for $U=0$(dotted line), $U=10^{-4}$,  $U=5\times10^{-4}$
 $U=10^{-3}$, $U=5\times10^{-3}$, and $U=10^{-2}$(solid line).}
For vanishing interactions $U$ (dotted line) all the particles are
transported once the cycle encloses the value $X_2=-1$
(mega-crossing). As $U$ is being increased the transport gradually
changes leading eventually to a step-like behavior (solid line) for large
repulsive interactions (sequential crossing).}
 
\end{figure}

\section{Transport during stirring:\,\,\,\,\, the~case~${\mathbf U \ne 0}$}

If we have a rectangular pumping cycle in $X$-space, it can be closed
at $X_{2}=\pm\infty$, where the influence of the change in $X_{1}$
can be safely neglected since the monopoles are located on the $X_{1}=0$
axis around $\varepsilon\approx-1$. Accordingly, 
the predominant contribution to $Q$ results from the $dX_{2}$ variation and
therefore we refer from now on to ${G(\varepsilon) \equiv G_{2}(X)}$ only. 
An overview of the numerical results for the conductance is shown 
in Figs.~(\ref{fig:E-n-G1}-\ref{fig:E-n-G2}),  
where we plot $G$ as a function of ${X_{2}=\varepsilon}$ 
for various interaction strengths $U$. 
Besides $G$ we also plot the $X_{2}$-dependence of the energy levels
and of the site population. Five representative values of $U$ are
considered including also the case of weak/strong attractive 
interactions $U<0$. 
Finally, in Fig.~\ref{fig:QvsR} we plot the implied dependence 
of $Q$ on the radius of the pumping cycle, again for 
several representative values of~$U$.

\subsection{Mega crossing }
For small positive values of $U$, the dynamics appears to be the same as in the $U=0$ case. Namely, 
all the particle cross ``together'' from the shuttle orbital to the $\varepsilon_{-}$ canal 
orbital. We call this type of dynamics ``mega crossing''. In the language of the adiabatic 
picture this means that all the Dirac monopoles are piled up in the center of the 
pumping cycle. Once the interaction~$U$ is turned on this ``pile" disintegrates into $N$ 
elementary ``monopoles". Consequently the dependence of $Q$ on the radius of the pumping 
cycle becomes of importance. Disregarding {\em fluctuations} that reflect discreteness of 
the magnetic charge, the value of~$Q$ is determined by the number of monopoles that are 
encircled by the pumping cycle. Thus, by measuring $Q$ versus $R$ we obtain information 
on the distribution of the monopoles and hence on the strength of the interatomic interactions.

\subsection{Sequential crossing }

In the sequential crossing regime the transport can be regarded 
as a sequence of LZ crossings. Each LZ crossing is formally 
the same as the LZ crossing of a single particle in a two-site 
system, while the topology is reflected by the splitting ratio $\lambda$ 
and the interaction is reflected in the scaled 
coupling constants $\kappa_n$. Accordingly we get    
\be{0}
G=-\frac{1}{2}\lambda
\sum_{n=1}^{N}\frac{2\kappa_{n}^{2}}{[(\varepsilon-\varepsilon_{n})^{2}+(2\kappa_{n})^{2}]^{3/2}}.
\label{eq:G-anal-sequential}
\ee
We overplot this formula in the lower panel of Fig.~\ref{fig:E-n-G1}c 
where an excellent agreement is observed.

\subsection{Gradual crossing}

For intermediate values
of $U$ (weak repulsive interaction), i.e. in the range $\kappa/N \ll U\ll N\kappa $,
we find neither the sequential crossing of Eq.~(\ref{eq:G-anal-sequential}),
nor the mega-crossing of Eq.~(\ref{eq:G-anal-mega}), but rather
a gradual crossing. Namely, in this regime, over a range $\Delta X_{2}=(3/2)(N{-}1)U$
we get a constant geometric conductance:  
\be{0}
G\,\,\,\,\approx\,\,\,\,-\lambda\frac{1}{3U}
\label{eq:G-anal-intermediate}
\ee
which reflects in a simple way the strength of the interaction. This
formula has been deduced by extrapolating Eq.~(\ref{eq:G-anal-sequential}),
and then was validated numerically (see lower panel of Fig.~\ref{fig:E-n-G1}b).
In this regime it is more illuminating to plot the scaled conductance 
\be{13}
\hat{G}(\hat{\varepsilon}) \ \ \equiv \ \ \left[\frac{Q}{(N{-}1)U}\right]^{-1} \, G(\varepsilon)
\ee
which is implied by the scaling (\ref{e12}) of $\varepsilon$.
The numerical results are reported in Fig.~\ref{fig:G-scaled}. 
The shape of the plot depends only on the dimensionless 
parameters $U/\kappa$ and $N$. The curves corresponding 
to different $\lambda$ and $U$ values 
(but with the same constant ratio $U/\kappa$) 
fall nicely one onto the other with good accuracy, 
confirming the scaling relation (\ref{e13}).

\subsection{Attractive interaction}

So far we have discussed repulsive interactions
for which the $N$-fold ``degeneracy'' of the $U=0$ 
mega crossing is lifted and we get a sequence 
of $N$~avoided crossings. Also for $U<0$ 
this $N$-fold ``degeneracy'' is lifted, but in a different
way: The levels separate in the ``vertical'' (energy) direction
(see upper panels of Fig.~\ref{fig:E-n-G2}) rather than 
``horizontally'' (see upper panels of Fig.~\ref{fig:E-n-G1}).

In the $U<0$ regime all the particles execute a single two-level
transition from the shuttle to the canal (see Fig.~\ref{fig:E-n-G2}a).
This transition happens directly from the ${n=N}$ state to the ${n=0}$ 
state. The coupling between these two states is exponentially     
small in $N$ because it requires a virtual $N$-th order transition 
in perturbation theory via the intermediate $n$ values.  

For sufficiently strong attractive interactions (${|NU|\gg 1}$)
all the particles are glued together and behave like a classical ball
that rolls from the shuttle to \emph{one of the canal sites} (see
Fig.~\ref{fig:E-n-G2}b). When the sign of $X_{1}$ is reversed
the ball rolls from one end of the canal to the other end (not shown).
This has to be clearly distinguished from the $N$-fold degenerated
transition to the \emph{lower canal level} 
which is observed in the $U{=}0$ case.


\section{Conclusions}

In this paper we have considered the transport induced in
a few site system that contains condensed 
particles. One new aspect that has emerged throughout the 
analysis is the existence of four distinct dynamical regimes 
depending on the strength of the interatomic interactions. 
This observation also applies to studies of LZ crossings 
in dimer systems and has further implications regarding the
quantum stirring in topologically non-trivial systems.    
It should be clear that the analysis of our ``driven vortex'' 
requires the toolbox of adiabatic processes,  
and it should be distinguished from the ignited
stirring of Refs.~\cite{RK99,MCWD00}. 

The actual measurement of induced
neutral currents poses a challenge to experimentalists. In fact, there
is a variety of techniques that have been proposed for this purpose.
For example one can exploit the Doppler effect at the perpendicular
direction, which is known as the rotational frequency shift \cite{doppler}.

The analysis of the prototype trimer system reveals the crucial importance
of interactions. The interactions are not merely a perturbation but
determine the nature of the transport process. We expect the induced
circulating atomic current to be extremely accurate, which would open
the way to various applications, either as a new metrological standard,
or as a component of a new type of quantum information or processing
device.

\begin{acknowledgments}
This research was supported by a grant from the United States-Israel
Binational Science Foundation (BSF).
\end{acknowledgments}


\appendix


\section{The Berry phase and the~$\mathbf{B}$-field}

In this Appendix we explain how the field $\mathbf{B}$ 
emerges in the theory of Berry phase. 
In the presentation below we follow the notations as in \cite{pmc}. 
Given a time dependent Hamiltonian $\mathcal{H}(X(t))$ 
where ${X=(X_1,X_2,X_3)}$ are control parameters 
it is convenient to expand the evolving wavefunction 
in the adiabatic basis   
\be{0}
\Big| \Psi(t) \Big\rangle = \sum_{n}a_{n}(t) \Big| n(X(t)) \Big\rangle.
\ee
Then the Schr\"{o}dinger equation $i \partial_{t}\ket{\psi}=\mcal{H}(X(t))\ket{\psi}$
becomes 
\be{0}
\frac{da_n}{dt} = 
-i E_{n}a_{n}+i\sum_{m}\sum_{j}\dot{X}_{j}A_{nm}^{j}a_{m},
\label{app2-eq:adiabatic-equation}
\ee
where we defined   
\be{0}
A_{nm}^{j} = i \Big\langle m \Big| \frac{\partial}{\partial X_{j}} n \Big\rangle.
\label{eq:appA1}
\ee
Differentiation by parts of $\partial_{j}\langle m(X)|n(X)\rangle=0$ implies
that $A_{nm}^{j}$ is a Hermitian matrix. We denote its (real)
diagonal elements as 
\be{0}
\mathbf{A}^{j}(X) \ \ \equiv \ \ A_{nn}^{j}.
\ee
The line integral over the vector field $\mathbf{A}(X)$ along  
a closed driving cycle gives the Berry phase
\be{0}
\mbox{BerryPhase} = \oint  \mathbf{A} \cdot dX.
\ee
Using $\partial_{X_{j}}\braket{m(X)}{\mcal{H}}{n(X)}=0$ we find
that the off-diagonal elements of $\mathbf{A}_{nm}^{j}$ can be written as 
\be{0}
A_{nm}^{j}
=\frac{i}{E_{m}{-}E_{n}} 
\Big\langle n \Big| \frac{\partial\mcal{H}}{\partial X_{j}} \Big| m \Big\rangle
=\frac{-i\mcal{F}_{nm}^{j}}{E_{m}{-}E_{n}}.
\ee
The ``1-form" $\mathbf{A}^{j}$ is formally like a vector potential, 
and we can associate with it a gauge invariant ``2-form" $B^{kj}$ 
which is formally like a magnetic field:
\be{0}
B^{kj} & = & \partial_{k}\mathbf{A}^{j}-\partial_{j}\mathbf{A}^{k}
\nonumber \\
 &=& -2 \mbox{Im}\scap{\partial_{k}n}{\partial_{j}n}
\nonumber \\
 &=& -2 \mbox{Im}\sum_{m}\mathbf{A}_{nm}^{k}\mathbf{A}_{mn}^{j}
\nonumber\\
 &=& \sum_{m\ne n}\frac{2 \,\textrm{Im}[\mathcal{F}_{nm}^{k}\mathcal{F}_{mn}^{j}]}{(E_{m}{-}E_{n})^{2}}.
\label{app2-eq:G-general-general}
\ee
In order to make the magnetic field analogy more 
transparent we assume that we have three control  
parameters $(X_1,X_2,X_3)$. Then it is natural 
to associate with the antisymmetric matrix $B^{kj}$ 
a field whose components are $\mathbf{B}_1=-B^{32}$ 
and $\mathbf{B}_2=B^{31}$ and $\mathbf{B}_3=B^{12}$.  
This $\mathbf{B}(X)$ field has has zero divergence 
everywhere with the exception of the Dirac monopoles 
at points of degeneracy. Dirac monopoles 
have quantized charge such that their flux 
is an integer multiple of $2\pi$. The Dirac 
monopoles must be quantized like that,   
otherwise Stokes' theorem would imply 
that the Berry phase is ill-defined.


\section{The Kubo formula and the~$\mathbf{B}$-field}

The Kubo formula is traditionally used in order to calculate 
the response of a driven system in the linear response regime. 
Given a time dependent Hamiltonian $\mathcal{H}(X(t))$ 
the linear DC-response of the system is expressed as
\be{0}
\langle \mathcal{F}^k \rangle = -\sum_{j} G^{kj} \dot{X}_{j},
\label{eq:appB}
\ee
where the generalized conductance matrix~$G^{kj}$ 
can be calculated from the Kubo formula.
This matrix can be decomposed in a symmetric and
an anti-symmetric part which account 
for the dissipative and non-dissipative effect 
of the driving respectively. 
Here we consider a strictly adiabatic
driving: Although the energy is not a constant 
of motion the system returns to the initial state 
at the end of each cycle. In this case
there is no dissipation and thus we consider only 
the anti-symmetric (``geometric") part of $G^{kj}$ 
which is identified as $B^{kj}$.  
We further illuminate this identification in the next appendix.

In the stirring problem there are two control parameters 
which we call $X_1$ and $X_2$, and the current 
operator $\mathcal{I}=\mathcal{F}^{3}$
is conveniently regarded as conjugate 
to a fictitious Aharonov-Bohm flux parameter $X_{3}$. 
If the particles were charged we could 
regard $X_3$ as an actual control parameter;
then $\dot{X}_3$ would be the electro-motive-force
and $G^{33}$ would be the conventional Ohmic conductance.
In Section~4 we use simplified indexing, 
namely $G_j=G^{3j}$. Accordingly Eq.(\ref{eq:appB}) leads 
to Eq.(\ref{e3}) and in the adiabatic limit 
we obtain  Eq.(\ref{e7}) with Eq.(\ref{e8}) 
which follows from Eq.(\ref{app2-eq:G-general-general}).


\section{Alternative derivation for the geometric conductance}

The definition of $\mathbf{B}$ in Appendix~A illuminates 
its geometric interpretation in the context of the 
the Berry phase formalism, but does not help in understanding  
why it emerges in the linear response calculation 
as described in Appendix~B.   
For this reason it might be helpful to derive  Eq.(\ref{Ggeom1}) 
directly from  the Schr\"odinger equation. 
For the LZ crossing problem it becomes 
\be{0}
\frac{d{a}_{n}}{dt} =  -i E_{n}a_{n}+ i\dot{\varepsilon}\sum_{m}\mathbf{A}_{nm}a_{m},
\ee
where $\dot{\varepsilon}$ is a small parameter.
Using the explicit expressions Eq.(\ref{e74}) for the 
adiabatic eigenstates, the definition Eq.(\ref{eq:appA1}),
and the identities   
${\partial_x\arctan(x)=1/(1{+}x^{2})}$ 
and ${\partial\theta/\partial\varepsilon=2\kappa/\Omega^{2}}$
we find
\be{0}
A_{mn}=\frac{\kappa}{\Omega^{2}}
\left(
\amatrix{
0 & -i\cr
i & 0}
\right)
\ee
The zero order adiabatic eigenstates of 
the Hamiltonian $\tilde{\mathcal{H}}=\mathcal{H}-\dot{\varepsilon}\mathbf{A}$ 
are $|E_{-}\rangle$ and  $|E_{+}\rangle$ of Eq.(\ref{e74}). 
The first order adiabatic eigenstates are found 
from perturbation theory. In particular the lower state is  
\be{0}
|-\rangle & = & |E_{-}\rangle-i\dot{\varepsilon}\frac{\kappa}{\Omega^{3}}|E_{+}\rangle.
\ee
It is important to realize that the expectation value 
of the current operator vanishes in zeroth order. 
This is because the the zeroth order Hamiltonian $\tilde{\mathcal{H}}$ 
is time-reversal symmetric. It is the $\dot{\varepsilon}$  perturbation 
term that breaks the time-reversal symmetry leading to 
\be{0}
\langle - \mid\mathcal{I}\mid -\rangle 
= 2\dot{\varepsilon} \frac{\kappa^{2}}{\Omega^{3}}.
\label{app2-eq:I-expectation-value}
\ee
Using the notation $\langle I\rangle=-G\dot{\varepsilon}$ 
we see that this result is in agreement with Eq.(\ref{Ggeom1}) as expected.




\begin{thebibliography}{}


\bibitem{AK98}
B.P. Anderson et al. \hide{M.A. Kasevich}, Science {\bf 282}, 1686 (1998). 
D. Jaksch et al., Phys. Rev. Lett. {\bf 81}, 3108 (1998);
C. Orzel et al., Science {\bf 291}, 2386 (2001);
M. Greiner et al., Nature {\bf 415}, 39 (2002);
I. Bloch, Nature Phys. {\bf 1}, 23-30 (2005);
R. Folman et al., Phys. Rev. Lett. 84, 4749  (2000).

\bibitem{HHHR01} 
W. H\"ansel, et al., Nature {\bf 413}, 498 (2001); 
P. Hommelhoff, et al., New J. Phys. {\bf 7}, 3 (2005).

\bibitem{MDJZ04}A. Micheli, et al., Phys. Rev. Lett. {\bf 93}, 140408 (2004);
B. T. Seaman, et al., (2006) [cond-mat/0606625].

\bibitem{SAZ07} 
J. A. Stickney, D. Z. Anderson, A. A. Zozulya, 
Phys. Rev. A {\bf 75}, 013608 (2007).

\bibitem{SFC02} 
J. Schmiedmayer, R. Folman, T. Calarco, J. Mod. Opt. {\bf 49}, 1375 (2002).

\bibitem{SHAWGBSK05} T. Schumm, et al., Nature 1, {\bf 57} (2005).;
Y-J Wang, et al., Phys. Rev. Lett. {\bf 94}, 090405 (2005);
E. Andersson, et al., Phys. Rev. Lett. {\bf 88}, 100401 (2002);
M. R. Andrews, et al., Science {\bf 275}, 637 (1997).

\bibitem{MAKDTK97} M. O. Mewes, et al., Phys. Rev. Lett. {\bf 78}, 582 (1997); E. W. Hagley, et al.,
Science {\bf 283}, 1706 (1999); Y. Shin, et al., Phys. Rev. Lett. {\bf 92}, 050405 (2004).

\bibitem{SMSKE04}T. Stoferle, H. Moritz, C. Schori, M. Kohl, T. Esslinger, Phys. Rev. Lett. {\bf 92}, 130403 (2004); 
C. Schori, T. Stoferle, H. Moritz, M. Kohl, T. Esslinger, Phys. Rev. Lett. {\bf 93}, 240402 (2004).

\bibitem{KIGHS06} C. Kollath, A. Iucci, T. Giamarchi, W. Hofstetter, U. Schollw\"ock,
Phys. Rev. Lett. {\bf 97}, 050402 (2006); A. Iucci, 
M. Cazalilla, A. Ho, T. Giamarchi, Phys. Rev. A {\bf 73}, 041608(R) (2006); A.M. Rey, P. B. Blakie, G. 
Pupillo, C. Williams, C. W. Clark, Phys. Rev. A {\bf 72}, 023407 (2005); G. G. Batrouni, F. F. Assaad,
R. T. Scalettar, P. J. H. Denteneer, Phys. Rev. A {\bf 72}, 031601(R) (2005); E. Lundh, 
Phys. Rev. A {\bf 70}, 061602(R) (2004).

\bibitem{JM06}M. J\"a\"askel\"ainen, P. Meystre, Phys. Rev A {\bf 73}, 013602 (2006);
D. R. Dounas-Frazer, L. D. Carr, quant-ph/0610166 (2006); K. W. Mahmud, H. Perry, W. P.
Reinhardt, J. Phys. B {\bf 36}, L265 (2003); M. Albiez et al., Phys. Rev. Lett. {\bf 95}, 
010402 (2005). 

\bibitem{niu} B. Wu and J. Liu, Phys. Rev. Lett. {\bf 96}, 020405 (2006);
J. Liu, B. Wu, Q. Niu, Phys. Rev. Lett. {\bf 90}, 170404 (2003).

\bibitem{raizen}C.-S. Chuu, et. al, Phys. Rev. Lett. {\bf 95}, 260403 (2005);
A. M. Dudarev, M. G. Raizen, and Q. Niu, Phys. Rev. Lett. {\bf 98}, 063001 (2007).

\bibitem{AOC05} 
L. Amico, A. Osterloh, F. Cataliotti, 
Phys. Rev. Lett. {\bf 95}, 063201 (2005).


\bibitem{pumping} 
D. J. Thouless,
Phys. Rev. B {\bf 27} 6083 (1983).
Q. Niu and D. J. Thouless, J. Phys. A {\bf 17} 2453 (1984).
M. Buttiker, H. Thomas and A Pretre, 
Z.~Phys.~B-Condens.~Mat., {\bf 94}, 133-137 (1994).  
P. W. Brouwer, 
Phys. Rev. B {\bf 58}, R10135 (1998).
B. L. Altshuler, L. I. Glazman, 
Science {\bf 283}, 1864 (1999).
M. Switkes, et al.,
Science {\bf 283}, 1905 (1999).

\bibitem{stirring} G. Rosenberg and D. Cohen, J. Phys. A {\bf 39}, 2287 (2006), 
and further references therein.

\bibitem{HKC08a} 
M. Hiller, T. Kottos and D. Cohen, Europhys. Lett. {\bf 82}, 40006 (2008).


\bibitem{pmc}
D. Cohen, 
Phys. Rev. B {\bf 68}, 155303 (2003).

\bibitem{Berry}
M.V. Berry, 
Proc. R. Soc. Lond. A {\bf 392}, 45 (1984).
J.E. Avron, A. Raveh and B. Zur, 
Rev. Mod. Phys. {\bf 60}, 873 (1988).
M.V. Berry and J.M. Robbins, 
Proc. R. Soc. Lond. A {\bf 442}, 659 (1993).



\bibitem{HKG06} 
M. Hiller, T. Kottos, and T. Geisel, 
Phys. Rev. A {\bf 73}, 061604(R) (2006), 
and further references therein.

\bibitem{BHK07} 
J. D. Bodyfelt, M. Hiller, and T. Kottos, 
Europhys. Lett. {\bf 78}, 50003 (2007).

\bibitem{FP03}R. Franzosi, V. Penna, Phys. Rev. E {\bf 67}, 046227 (2003);
K. Nemoto, et al., Phys. Rev. A {\bf 63}, 013604 (2000).
 



\bibitem{cnb} M. Chuchem and D. Cohen, J. Phys. A {\bf 41}, 075302 (2008).



\bibitem{MCWD00}K. W. Madison, et al.,
Phys. Rev. Lett. 84, 806 (2000).

\bibitem{RK99} C. Raman et. al., 
Phys. Rev. Lett. {\bf 83}, 2502 (1999). 

\bibitem{doppler}
I. Bialynicki-Birula and Z. Bialynicka-Birula, 
Phys. Rev. Lett. {\bf 78}, 2539 (1997).


\end{thebibliography}
\end{document}